\documentclass[12pt]{iopart}
\usepackage{graphicx}
\usepackage{multirow}
\usepackage{dcolumn}

\begin{document}

\title[Nonhermitian oscillators]{On the eigenvalues of some nonhermitian oscillators}
\author{Francisco M Fern\'andez \ and Javier Garcia}

\address{INIFTA (UNLP, CCT La Plata-CONICET), Divisi\'on Qu\'imica Te\'orica,
Blvd. 113 S/N,  Sucursal 4, Casilla de Correo 16, 1900 La Plata,
Argentina}

\ead{fernande@quimica.unlp.edu.ar}

\maketitle

\begin{abstract}
We consider a class of one-dimensional nonhermitian oscillators
and discuss the relationship between the real eigenvalues of
PT-symmetric oscillators and the resonances obtained by different
authors. We also show the relationship between the strong-coupling
expansions for the eigenvalues of those oscillators. Comparison of
the results of the complex rotation and the Riccati-Pad\'{e}
methods reveals that the optimal rotation angle converts the
oscillator into either a PT-symmetric or an Hermitian one. In
addition to the real positive eigenvalues the PT-symmetric
oscillators exhibit real positive resonances under different
boundary conditions. They can be calculated by means of the
straightforward diagonalization method. The Riccati-Pad\'e method
yields not only the resonances of the nonhermitian oscillators but
also the eigenvalues of the PT-symmetric ones.
\end{abstract}

\section{Introduction}

\label{sec:intro}

In a recent paper Jentschura et al\cite{JSLZJ08} discussed the
resonances for the anharmonic oscillator
$H=-\frac{1}{2}\frac{d^{2}}{dq^{2}}+\frac{1}{2}q^{2}+\sqrt{g}q^{3}$and
their weak- and strong-coupling expansions. They showed analytical
expressions for the coefficients of the former and numerical
estimates for those of the latter. In particular, the leading
coefficients of the strong-coupling expansions are the eigenvalues
of $H=-\frac{1}{2}\frac{d^{2}}{dq^{2}}+q^{3}$.

Some time earlier Bender and Boettcher\cite{BB98} had discussed
the eigenvalues of PT-symmetric oscillators of the form
$H=-\frac{d^{2}}{dx^{2}} -(ix)^{N}$ that exhibit a finite number
of real positive eigenvalues for $1<N<2$ and an infinite number
when $N\geq 2$.

Alvarez\cite{A95} discussed the analytical properties of the
solutions of the Hamiltonian operator
$H=\frac{1}{2}p^{2}+\frac{1}{2}kx^{2}+gx^{3}$ and showed that it
supports real and complex resonances depending on the complex
values of the coupling constant $g$. His results suggest that the
resonances calculated by Jentschura et al\cite{JSLZJ08} and the
real eigenvalues obtained by Boettcher and Bender\cite{BB98} (see
also Bender\cite{B07}) may by related in a simple way by means of
the Symanzik scaling\cite{A88} already invoked by Alvarez is his
investigation\cite{A95}. In exactly the same way the
strong-coupling expansion obtained by Jentschura et
al\cite{JSLZJ08} may be related to that obtained some time earlier
by Fern\'{a}ndez et al\cite{FGRZ98} for the PT-symmetric
oscillator $H=p^{2}+ix^{3}+\lambda x^{2}$. The purpose of this
paper is the exploration into such relationships as well as into
other properties of a class of nonhermitian oscillators.

In section~\ref{sec:eigenvalues} we investigate the relationship
among some of the earlier results on one-dimensional nonhermitian
oscillators. In section~\ref{sec:RPM} we discuss the application
of the complex-rotation\cite {YBLBF78} and
Riccati-Pad\'{e}\cite{FMT89a,FT96} methods to those oscillators.
Finally, in section~\ref{sec:conclusions} we summarize the main
results and draw conclusions.

\section{Real and complex eigenvalues}

\label{sec:eigenvalues}

As outlined above, Jentschura et al\cite{JSLZJ08} discussed several
properties of the resonances for the anharmonic oscillator
\begin{equation}
H_{c}=-\frac{1}{2}\frac{d^{2}}{dq^{2}}+\frac{1}{2}q^{2}+\sqrt{g}q^{3}
\label{eq:H_c}
\end{equation}
as well as their weak-coupling
\begin{equation}
E_{n}(g)=\sum_{k=0}^{\infty }E_{n,k}g^{k}  \label{eq:series_weak}
\end{equation}
and strong-coupling expansions
\begin{equation}
E_{n}(g)=g^{1/5}\sum_{k=0}^{\infty }L_{n,k}g^{-2k/5}
\label{eq:series_strong}
\end{equation}
The coefficients of the former can be obtained exactly by means of
perturbation theory and those of the latter in a numerical way. In
particular, the leading coefficients of the strong-coupling
expansions $L_{n,0}$ are the eigenvalues of the pure cubic
anharmonic oscillator
\begin{equation}
H_{l}=-\frac{1}{2}\frac{d^{2}}{dq^{2}}+q^{3}  \label{eq:H_l}
\end{equation}

On the other hand, the closely related PT-symmetric oscillators
\begin{equation}
H_{PT}=-\frac{d^{2}}{dx^{2}}-(ix)^{N}  \label{eq:H_PT}
\end{equation}
exhibit an infinite number of real positive eigenvalues when
$N\geq 2$\cite{BB98} and accurate results for $N=3$ and $N=4$ are
available for comparison\cite{B07}.

It is not difficult to obtain a connection between the results outlined
above by means of the Symanzik scaling
\begin{equation}
U^{\dagger }pU=\gamma ^{-1}p,\;U^{\dagger }xU=\gamma x  \label{eq:Symanzik}
\end{equation}
where $U$ is a well known unitary operator\cite{A88}. This
transformation was already used by Alvarez in his investigation of
the cubic anharmonic oscillator\cite {A95}. For example, if we
take into account that $2\gamma ^{2}U^{\dagger }H_{l}U=H_{PT}$
when $\gamma =(i/2)^{1/5}$ then we realize that the complex
eigenvalues $L_{n,0}$ of $H_{l}$ calculated by Jentschura et
al\cite{JSLZJ08} and the real positive eigenvalues $E_{n}^{PT}$ of
$H_{PT}$ for $N=3$ calculated by Boettcher and Bender\cite{BB98}
and Bender\cite{B07} are related by
\begin{equation}
L_{n,0}=2^{-3/5}i^{-2/5}E_{n}^{PT}  \label{eq:BB->J}
\end{equation}

Some time ago Fern\'{a}ndez et al\cite{FGRZ98} obtained the perturbation
expansion for the PT-symmetric oscillator
\begin{equation}
H_{F}=p^{2}+ix^{3}+\lambda x^{2}  \label{eq:H_F}
\end{equation}
in the form
\begin{equation}
E_{n}(\lambda )=\sum_{j=0}^{\infty }W_{n,j}\lambda ^{j}
\label{eq:series_lambda}
\end{equation}
Arguing as before, we can obtain the coefficients of the
strong-coupling expansion (\ref{eq:series_strong}) from those of
the perturbation series (\ref{eq:series_lambda}) as follows:
\begin{equation}
L_{n,j}=2^{-(4j+3)/5}i^{(4j-2)/5}W_{n,j}
\end{equation}
The first coefficients are shown in Table~\ref{tab:strongser} as an
illustrative example.

\section{Complex-rotation and Riccati-Pad\'{e} methods}

\label{sec:RPM}

For concreteness we consider the family of anharmonic oscillators
\begin{eqnarray}
H_{K} &=&\frac{1}{2}p^{2}-x^{K},  \nonumber \\
H_{K}\psi &=&E\psi  \label{eq:H_K}
\end{eqnarray}
The complex-rotation method (CRM) consists of the diagonalization of the
rotated Hamiltonian operator
\begin{equation}
U^{\dagger }H_{K}U=\gamma ^{-2}\left( \frac{1}{2}p^{2}-\gamma
^{K+2}x^{K}\right)  \label{eq:U+H_KU}
\end{equation}
where $\gamma =\eta e^{i\theta }$. The parameter $\eta >0$
produces a dilatation or contraction of the scale and $\theta $ a
rotation of the coordinate in the complex $x$-plane. On tuning
$\eta $ we improve the rate of convergence of the diagonalization
method as the matrix dimension increases and the value of $\theta
$ enables us to uncover the resonances\cite{YBLBF78}. For the
diagonalization method we choose the basis set of eigenfunctions
of the harmonic oscillator $H=p^{2}+x^{2}$.

For comparison purposes we also apply the Riccati-Pad\'{e} method
(RPM) for asymmetric potentials\cite{FT96}. It consists of the
expansion of the logarithmic
derivative of the eigenfunction $\psi (x)$%
\begin{equation}
f(x)=-\frac{\psi ^{\prime }(x)}{\psi (x)}
\end{equation}
in a Taylor series about the origin
\begin{equation}
f(x)=\sum_{j=0}^{\infty }f_{j}x^{j}
\end{equation}
where the coefficients $f_{j}$ depend on the two unknowns $E$ and
$f_{0}=-\psi ^{\prime }(0)/\psi (0)$. From the coefficients of the
even and odd powers of the coordinate $f_{e,j}=f_{2j}$ and
$f_{o,j}=f_{2j-1}$, $j=1,2,\ldots $, respectively, we construct
the Hankel determinants $H_{D}^{ed}(E,f_{0})=\left|
f_{e,i+j+d-1}\right| _{i,j=1}^{D}$, $H_{D}^{od}(E,f_{0})=\left|
f_{o,i+j+d-1}\right| _{i,j=1}^{D}$ and obtain both $E$ and $f_{0}$
from the roots of the system of nonlinear equations
$\{H_{D}^{ed}(E,f_{0})=0,\;H_{D}^{od}(E,f_{0})=0\}$. For every
fixed value of $d=0,1,\ldots $ we look for convergent sequences of
roots $E^{[D,d]}$, $D=2,3,\ldots $. Commonly, we obtain reasonable
results for $d=0$ but calculations with other values of $d$ enable
us to test the consistency of the method.

\begin{figure}[H]
\begin{center}
\includegraphics[width=9cm]{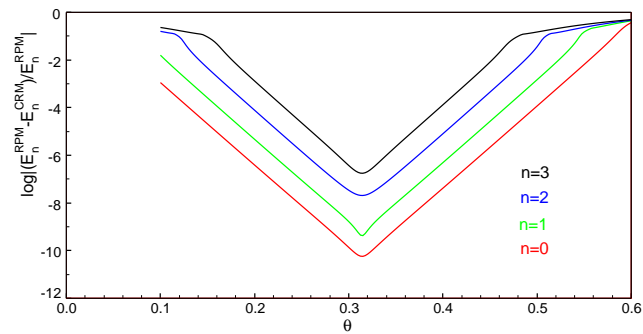}
\end{center}
\caption{$\log\left|\left(E_n^{RPM}-E_n^{CRM}\right)/E_n^{RPM}\right|$ for
the first resonances of the cubic oscillator. }
\label{fig:Loger}
\end{figure}

Since the rate of convergence of the RPM is considerably greater
than the one for the CRM we choose the results of the former as
exact or reference eigenvalues. Figure~\ref{fig:Loger} shows $\log
\left| \left( E_{n}^{RPM}-E_{n}^{CRM}\right) /E_{n}^{RPM}\right| $
as a function of $\theta $ for the first resonances of the cubic
oscillator ($K=3$). Those results suggest that the minimum of the
logarithmic deviation appears at $\theta =\pi /10$ (in all our
calculations we have chosen $\eta =e^{-1}$ that provides a
reasonable rate of convergence). In order to understand this
empirical result we resort to the scaling transformation
(\ref{eq:U+H_KU}) for $K=3,5,\ldots $. It is clear that
$U^{\dagger }H_{K}U=\gamma _{j}^{-2}\left[ \frac{1}{2}p^{2}-\gamma
_{j}^{K+2}x^{K}\right] $ is proportional to the PT-symmetric
oscillator $\frac{1}{2}p^{2}-(-1)^{j}ix^{K}$ when $\gamma
_{j}=e^{(2j+1)i\pi /[2(K+2)]}$, $j=0,1,\ldots ,K+1$. For $K=3$ and
$j=0$ we obtain $\theta =\pi /10$ as suggested by
Figure~\ref{fig:Loger}. The obvious conclusion is that the optimal
rotation angle converts each of the anharmonic oscillators of this
particular class into a PT-symmetric one.

\begin{figure}[H]
\begin{center}
\includegraphics[width=9cm]{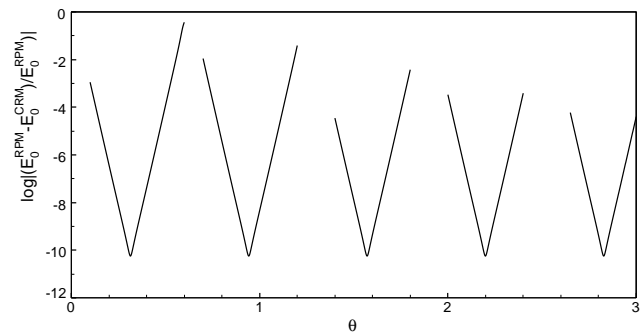}
\end{center}
\caption{Logarithmic error
$\log\left|\left(E_{0,j}^{RPM}-E_{0,j}^{CRM}
\right)/E_{0,j}^{RPM}\right|$ for the first set of eigenvalues of
the cubic oscillator ($K=3$) as functions of $\theta$. }
\label{fig:Loger2}
\end{figure}

It also follows from equation (\ref{eq:U+H_KU}) that $\gamma
_{j}^{2}U^{\dagger }H_{K}U=H_{K}$ when $\gamma _{j}=e^{2\pi ij/(K+2)}$.
Therefore, instead of just one eigenvalue $E_{n}$ we expect $K+1$ replicas
located at
\begin{equation}
E_{n,j}=e^{4\pi ij/(K+2)}E_{n},\;j=0,1,\ldots ,K+1  \label{eq::E_n,j}
\end{equation}
The RPM yields all these eigenvalues simultaneously as limits of
different sequences of roots of the same sequence of pairs of
Hankel determinants. On the other hand, the CRM uncovers them at
different values of $\theta $. Figure~\ref{fig:Loger2} shows $\log
\left| \left( E_{0,j}^{RPM}-E_{0,j}^{CRM}\right)
/E_{0,j}^{RPM}\right| $ as a function of $\theta $ for the lowest
resonance of the cubic oscillator. We appreciate that the closest
agreement between both methods takes place exactly at the rotation
angles $\theta _{j}=(2j+1)\pi /10$ derived above. Table~\ref
{tab:E0j} shows these results more precisely and
Table~\ref{tab:E0j_K5} a similar calculation for the quintic
oscillator.

\begin{figure}[H]
\begin{center}
\includegraphics[width=9cm]{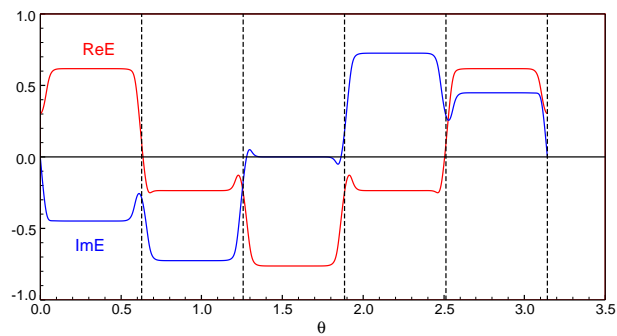}
\end{center}
\caption{Real and imaginary parts of the first resonance $E(\theta)$ for the
cubic oscillator ($K=3$). The vertical lines mark multiples of $\pi/5$. }
\label{fig:RR-K3}
\end{figure}

The case $j=0$ for the cubic oscillator agrees with the resonance
calculated by Jentschura et al\cite {JSLZJ08}. These authors
claimed to have chosen the rotation angle $\theta =\pi /5$ for all
their calculations on the cubic oscillator (in particular for the
strong-coupling expansion). However, we could not obtain
acceptable results for this rotation angle. In fact, our
calculations for the cubic oscillator suggest that the multiples
of $\theta =\pi /5$ are the worst choices. Figure~\ref{fig:RR-K3}
shows the real and imaginary parts of the first resonance as
functions of $\theta $. We appreciate that the regions of
stability appear at $j\pi /5<\theta <(j+1)\pi /5$, $j=0,1,2,3,4$
(the boundaries are marked by vertical dashed lines). The optimal
rotation angles discussed above (those that convert the anharmonic
oscillator into a PT-symmetric one) bisect each of these regions
and the rotation angle chosen by Jentschura et al corresponds to
one of the boundaries. Present results agree with those of
Alvarez\cite{A95} who proposed to integrate the differential
equation along the rays $\arg (\pm x)=\pi /10-\arg (g)/5$ in the
case of a harmonic oscillator perturbed by the cubic term
$gx^{3}$. More precisely, he also showed that the left and right
boundary conditions for the resonances hold in the common sector
$0<\frac{1}{2}\arg (g)+\frac{5}{2}\arg (x)<\frac{\pi }{2}$ so that
$0<\arg (x)<\frac{\pi }{5}$ for $g=1$ in agreement with the first
region of stability shown in Figure~\ref{fig:RR-K3}. The
appearance of the optimal rotation angle $\theta =\pi /5$ in the
paper by Jentschura et al\cite {JSLZJ08} is merely due to a
misprint\cite{J12}.

Table~\ref{tab:EnK3K5} shows the first resonances for the cubic and quintic
oscillators calculated by means of the RPM. They may be useful as benchmark
for testing other approximate methods. For example, the first three of them
for $K=3$ agree with those of Jentschura et al\cite{JSLZJ08}.

From the results just discussed one may be tempted to conclude
that the CMR with $\theta =0 $ should yield the eigenvalues of the
PT-symmetric oscillators. This conjecture is supported by the
convergence of this method towards the accurate RPM eigenvalues
shown in Table~\ref{tab:RR_ixN}. However, such conclusion is
wrong. Although the eigenvalues produced by two quite different
methods like the RPM and CRM agree accurately for all
$N=3,5,7\ldots $ only in the case $N=3$ they are those of the
PT-symmetric oscillators. For $N=5,7,\ldots $ both methods yield
the resonances discussed above rotated in the complex plane. In
fact, Bender and Boettcher\cite{BB97} clearly stated that the
diagonalization method is useful only for $1<N<4$ because in the
other cases the wedges in which the eigenfunction vanishes as
$|x|\rightarrow \infty $ do not contain the real $x$ axis. More
precisely, since those wedges are not symmetric about the origin
the complex rotation outlined above is insufficient to take into
account both the left and right PT boundary conditions\cite{BB98}.

Table~\ref{tab:compa_ixN} shows the first eigenvalues for the
PT-symmetric oscillators (\ref{eq:H_PT}) with $N=5$ and $N=7$
calculated by means of the RPM, CRM ($\theta =0$, $\eta =0.4$) and
WKB method. The first two approaches agree between them but not
with the WKB method that provides estimates to the actual
eigenvalues of the PT-symmetric oscillators\cite{BB98}. Note that
the discrepancy increases with the quantum number that makes the
WKB increasingly accurate. On the other hand, it is well known
that the eigenvalues of the Hamiltonian matrix agree with the WKB
ones for the $N=3$ case\cite{AF12}. As an additional confirmation
that the eigenvalues of the CRM are not those of the PT-symmetric
oscillators compare the results of Table~\ref{tab:compa_ixN} with
the accurate upper and lower bounds derived by Yan and
Handy\cite{YH01}. Although the functional form of the operators is
the same the boundary conditions are different\cite{BB98}. For
simplicity, from now we will refer to resonance\cite{A95} and
PT-symmetric boundary conditions\cite{BB98}. Although the CRM
takes into account only the former, it is interesting that it
yields real positive eigenvalues for the PT-symmetric oscillators
(\ref{eq:H_PT}) with $N=5,7,\ldots $. The reason is discussed
below.

\begin{figure}[H]
\begin{center}
\includegraphics[width=9cm]{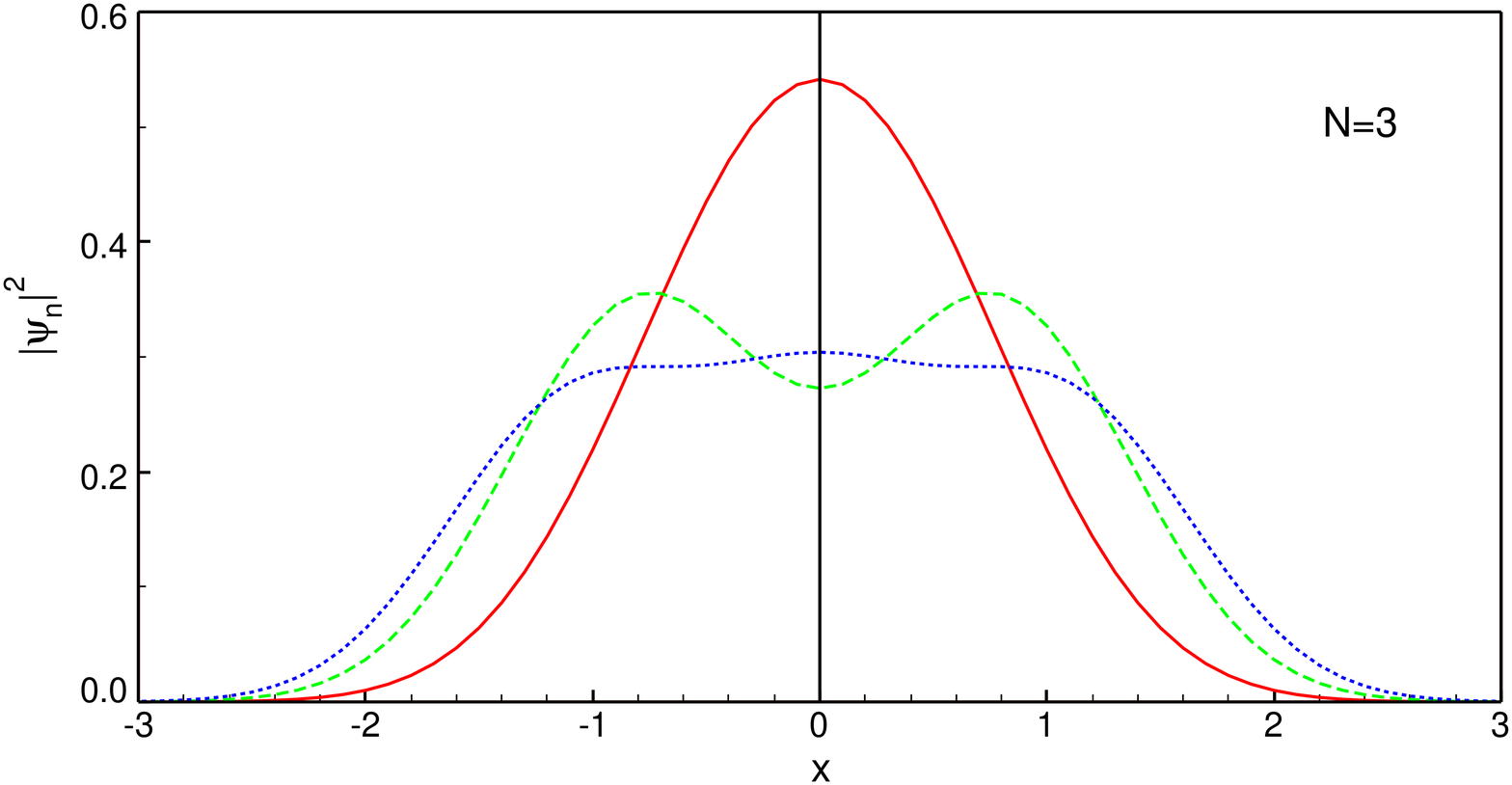} %
\includegraphics[width=9cm]{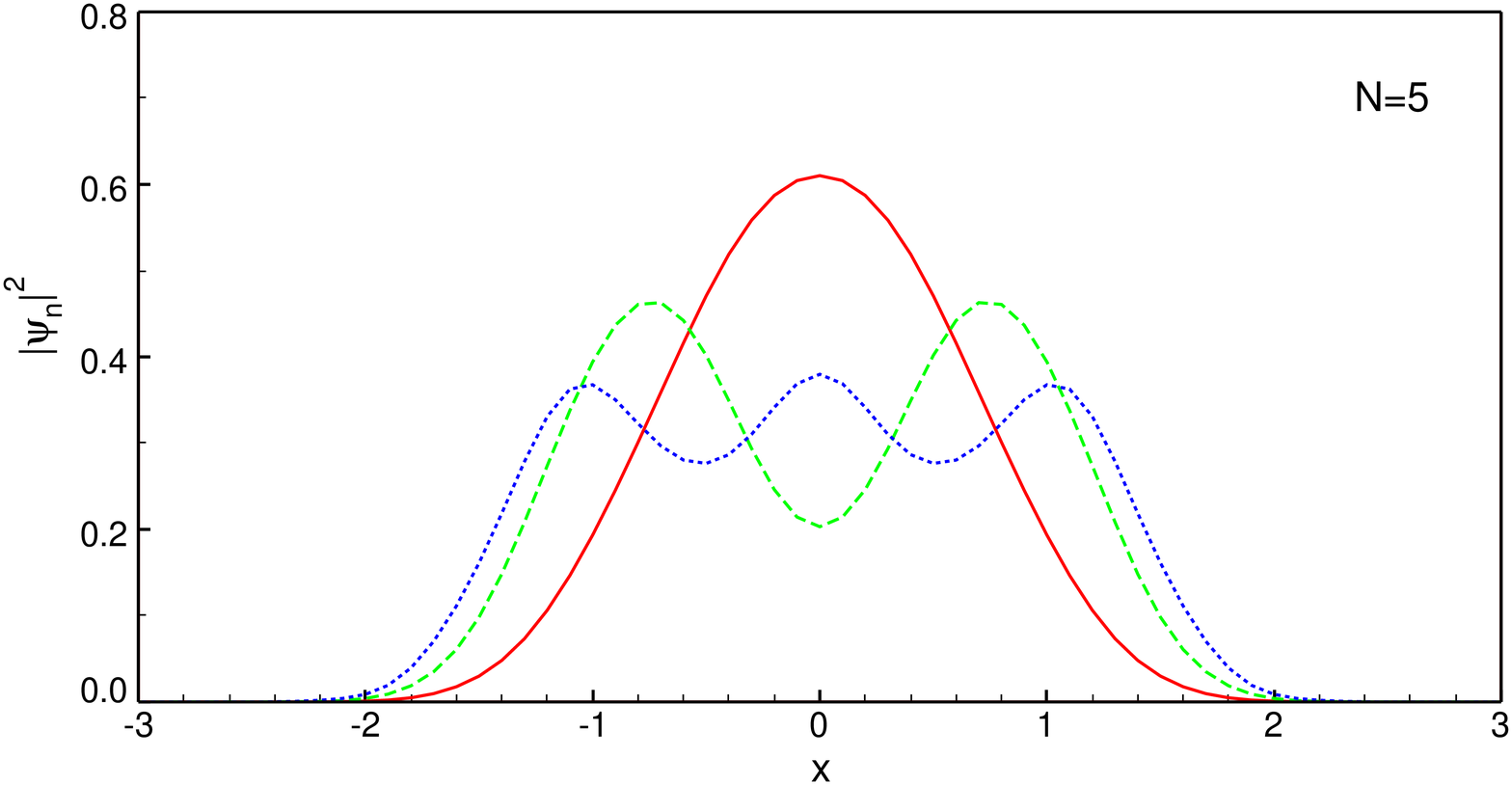} %
\includegraphics[width=9cm]{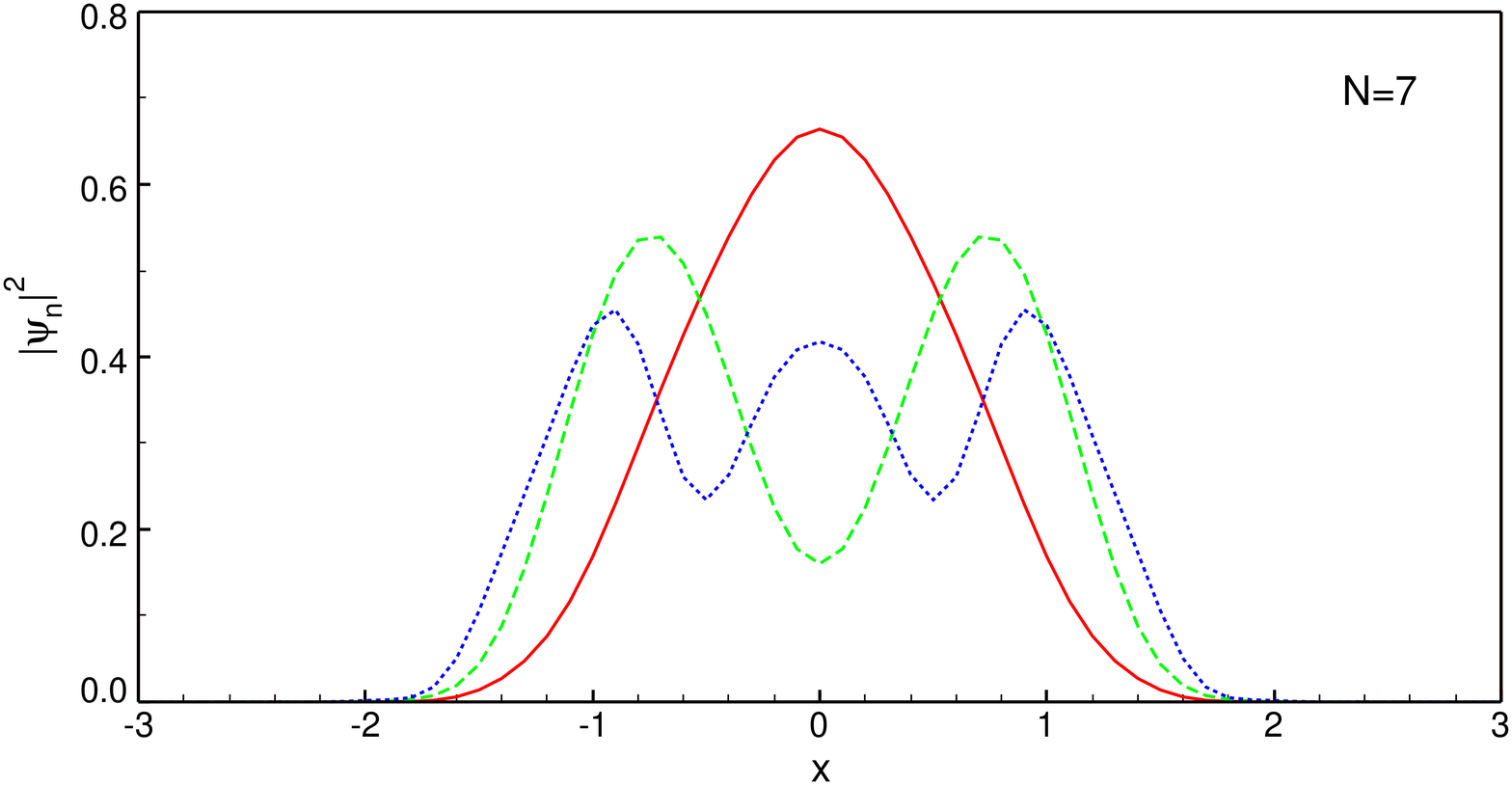}
\end{center}
\caption{$\left|\psi_n\right|^2$, $n=0$ (solid line, red), $n=1$ (dashed
line, green), $n=2$ (dotted line, blue) for the PT-symmetric oscillators (%
\ref{eq:H_PT}) with $N=3,5,7$ }
\label{fig:psi_PT}
\end{figure}

In order to understand the results just outlined we inspect the
form of the eigenfunctions provided by the CRM for the
PT-symmetric oscillators. We calculated the eigenfunctions $\psi
_{n}(x)$, $n=0,1,2$ and their absolute squares are shown in
Figure~\ref{fig:psi_PT} for $N=3,5,7$. We appreciate that they all
look similar and satisfy $\left| \psi _{n}(-x)\right| ^{2}=\left|
\psi _{n}(x)\right| ^{2}$as expected from the fact that $\psi
_{n}(-x)^{*}=\lambda \psi _{n}(x)$, where $|\lambda
|=1$\cite{B07}. More precisely, our numerical calculations suggest
that in these particular cases $\psi _{n}(-x)^{*}=(-1)^{n}\psi
_{n}(x)$. Even though the resonance boundary conditions are
different from the PT-symmetric ones for $N=5,7,\ldots $ there
appears to be an unbroken symmetry that produces real eigenvalues.
Besides, all those eigenfunctions are strongly localized about
$x=0$ as expected for a resonance. It is interesting that both the
RPM and the CRM yield real and positive eigenvalues with localized
eigenfunctions for the PT-symmetric oscillators although they are
not the true eigenvalues and eigenfunctions of the PT-symmetric
oscillators for $N>3$.

In addition to the resonances just discussed the RPM also yields
the true eigenvalues of the PT-symmetric oscillators for all
$N=3,5,\ldots $. For example, for $N=5$ we estimated
$E_{0}=1.9082646$ from determinants of dimension $D=10,\ldots
,20$. Note that this eigenvalue is considerably greater than that
in Table~\ref{tab:compa_ixN} obtained from the resonance boundary
condition. We will discuss this issue again below.

The situation is remarkably different for $K=4,6,\ldots $. The
PT-symmetric oscillators require asymmetric boundary
conditions\cite{BB98} and, consequently, one should apply the RPM
for asymmetric potentials outlined above. However, in the case of
resonances the boundary conditions are symmetric (for example,
outgoing waves to the right and left) and the oscillator exhibits
true even parity. In this case the appropriate logarithmic
derivative of the wavefunction is of the form
\begin{equation}
f(x)=\frac{s}{x}-\frac{\psi ^{\prime }(x)}{\psi (x)}
\end{equation}
where $s=0$ or $s=1$ for even or odd eigenfunctions, respectively. From the
coefficients of the Taylor expansion
\begin{equation}
f(x)=\sum_{j=0}^{\infty }f_{j}x^{2j+1}
\end{equation}
we construct the Hankel determinants $H_{D}^{d}(E)=\left| f_{i+j-1+d}\right|
_{i,j=1}^{D}$ that depend on the only unknown $E$ and obtain the eigenvalues
from sequences of roots of $H_{D}^{d}(E)=0$\cite{FMT89a}.

On the other hand, we can apply the CRM exactly in the same way discussed
above. In this case the optimal rotation angles are given by $\gamma
_{j}=e^{(2j+1)i\pi /(K+2)}$, $j=0,1,\ldots ,K+1$ that make $U^{\dagger
}HU=\gamma _{j}^{-2}\left( \frac{p^{2}}{2}+x^{K}\right) $ proportional to
the Hermitian operator $\frac{p^{2}}{2}+x^{K}$. Table~\ref{tab:E0j_K4}
compares the CRM and RPM results for the first resonance of the quartic
oscillator ($K=4$). There are $K/2+1$ replicas of every resonance given by
\begin{equation}
E_{n,j}=e^{-2\pi ij/(K/2+1)}E_{n},\;j=0,1,\ldots ,\frac{K}{2}
\end{equation}

In this case the RPM for asymmetric potentials yields the
eigenvalues of the corresponding PT-symmetric oscillator.
Table~\ref{tab:En_N4_N5} shows the first three eigenvalues of the
PT-symmetric oscillators (\ref{eq:H_PT}) with $N=4$ and $N=5$. The
results in the first column agree with those obtained earlier by
means of numerical integration\cite{BB98,B07} and those in the
second column lie within the upper and lower bounds derived by Yan
and Handy\cite{YH01}. The rate of convergence of the RPM is
considerably greater for $N=4$; in addition to it, we experienced
considerable numerical difficulties in obtaining the roots of the
pair of Hankel determinants for $N=5$ by means of the
Newton-Raphson method.

\section{Conclusions}

\label{sec:conclusions}

In section~\ref{sec:eigenvalues} we have shown that a simple
scaling argument enables one to connect the results obtained
earlier by several authors for a class of nonhermitian
oscillators. Although this relationship is contained in Alvarez's
work\cite{A95} the actual connection formulas have not been made
explicit as far as we know. For example, the resonances of the
cubic oscillator are straightforwardly related to the eigenvalues
of the corresponding PT-symmetric oscillator. Such connection is
not possible for other oscillators because the boundary conditions
that give rise to the resonances and PT-symmetric eigenvalues are
different.

The comparison of the RPM and CRM results enabled us to obtain the optimal
rotation angle for the latter approach. We have shown that the effect of the
optimal coordinate rotation is to convert the nonhermitian oscillators (\ref
{eq:H_K}) into either a PT-symmetric or Hermitian one, for $K$ odd or even,
respectively. Such results  are consistent with Alvarez's analysis of the
harmonic oscillator with a cubic perturbation\cite{A95}.

We have also shown that both the RPM and CRM yield real positive eigenvalues
for the PT-symmetric oscillators (\ref{eq:H_PT}) with $N=3,5,7,\ldots $ but
those results are not the actual eigenvalues of the PT-symmetric oscillators
when $N>3$ because the resonance and PT-symmetric boundary conditions are
different. The CRM eigenfunctions are strongly localized and their absolute
squares exhibit the symmetry coming from unbroken symmetry. It is also
interesting that the eigenfunctions for the case $N=3$ (where the CRM yields
the actual eigenvalues of the PT-symmetric oscillator) are similar to those
of $N=5,7$ (where the boundary conditions are those for the resonances).

On the other hand, the RPM yields both the real positive
resonances mentioned above for odd $N$ as well as the actual
PT-symmetric eigenvalues obtained by Bender and
Boettcher\cite{BB98} and Bender\cite{B07}. In the case of
$N=3,5,7,\ldots $ all the eigenvalues are limits of sequences of
roots of the same Hankel determinants given by the RPM for
nonsymmetric potentials. In the case of even-parity potentials
$N=4,6,\ldots $ the RPM for even parity potentials yields the
resonances and the approach for nonsymmetric potentials provides
the eigenvalues of the corresponding PT-symmetric oscillators. The
main disadvantage of this approach as a practical tool is that it
provides results for so many different problems that it is
sometimes difficult to pick up the correct sequence of roots of
the system of two Hankel determinants necessary for the treatment
of nonsymmetric problems. On the other hand, from a mathematical
point of view, this property of the RPM is most intriguing and
interesting.

\begin{table}[H]
\caption{Coefficients $L_{0,j}$ of the strong-coupling expansion}
\label{tab:strongser}
~(\ref{eq:series_strong})
\par
\begin{center}
\par
\begin{tabular}{|D{.}{.}{2}|D{.}{.}{25}|D{.}{.}{35}|}
\hline \multicolumn{1}{|c}{$j$} & \multicolumn{1}{|c}{Ref.\cite{JSLZJ08}} & \multicolumn{1}{|c|}{From Ref.\cite{FGRZ98}} \\
\hline

0&  0.617160050 - 0.448393023i &   0.617160049536-0.448393022571i        \\
1&  -0.013228193 + 0.040712191i&  -0.0132281928671+0.0407121914135i      \\
2&  0.009259259 + 0.000000000i &   0.0925925925868                       \\
3&  -0.000294361 - 0.000905951i&  -0.000294361224639-0.000905950695052i  \\
\hline

\end{tabular}
\end{center}
\end{table}

\begin{table}[H]
\caption{Lowest resonance for the cubic oscillator ($K=3$). The
first column shows the value of $j$ that determines the optimal
rotation angle $\theta_j=\frac{2j+1}{10}\pi$ for the CRM. The
first and second entries in the second and third columns
correspond to the CRM and RPM eigenvalues, respectively).}
\label{tab:E0j}
\begin{center}
\par
\begin{tabular}{|D{.}{.}{2}|D{.}{.}{20}|D{.}{.}{20}|}
\hline
\multicolumn{1}{|c}{$j$} & \multicolumn{1}{|c|}{$\Re\left (E\right )$} &
\multicolumn{1}{c|}{$\Im\left (E\right )$} \\ \hline
\multirow{2}{*}{0} & 0.6171600495373 & -0.448393022575 \\
& 0.61716004953893673754 & -0.44839302257593285633 \\ \hline
\multirow{2}{*}{1} & -0.2357341624247 & -0.725515150844 \\
& -0.23573416242530496269 & -0.72551515084615828994 \\ \hline
\multirow{2}{*}{2} & -0.762851774225 & 0.00000000000000 \\
& -0.76285177422726354970 & 0.00000000000000000000 \\ \hline
\multirow{2}{*}{3} & -0.2357341624247 & 0.725515150844 \\
& -0.23573416242530496269 & 0.72551515084615828994 \\ \hline
\multirow{2}{*}{4} & 0.617160049537 & 0.448393022575 \\
& 0.61716004953893673754 & 0.44839302257593285633 \\ \hline
\end{tabular}
\end{center}
\end{table}

\begin{table}[H]
\caption{Idem Table~\ref{tab:E0j} for the quintic oscillator
($K=5$); in this case the optimal rotation angle is
$\theta_j=\frac{2j+1}{14}\pi$.} \label{tab:E0j_K5}
\begin{center}
\begin{tabular}{|D{.}{.}{2}|D{.}{.}{20}|D{.}{.}{20}|}
\hline  \multicolumn{1}{|c}{$j$} &
        \multicolumn{1}{|c|}{$\Re\left (E\right )$} &
        \multicolumn{1}{c|}{$\Im\left (E\right )$} \\
\hline  \multirow{2}{*}{0}  & 0.639629797817        & -0.308029476062       \\
                & 0.63962979781725182920    & -0.30802947606177966696   \\
\hline \multirow{2}{*}{1}   & 0.1579755139908       & -0.692135950055       \\
                & 0.15797551399078843452    & -0.69213595005459668245   \\
\hline \multirow{2}{*}{2}   & -0.442637553984       & -0.555049936656       \\
                & -0.44263755398395529372   & -0.55504993665591387838   \\
\hline \multirow{2}{*}{3}   & -0.709935515648       & 0.00000000000000     \\
                & -0.70993551564816994002   & 0.00000000000000000000    \\
\hline  \multirow{2}{*}{4}  & -0.442637553984       & 0.555049936656        \\
                & -0.44263755398395529372   & 0.55504993665591387838    \\
\hline  \multirow{2}{*}{5}  & 0.1579755139909       & 0.692135950055        \\
                & 0.15797551399078843452    & 0.69213595005459668245    \\
\hline  \multirow{2}{*}{6}  & 0.639629797817        & 0.308029476062        \\
                & 0.63962979781725183008    & 0.30802947606177966739    \\
\hline
\end{tabular}
\end{center}
\end{table}

\begin{table}[H]
\caption{First resonances for the cubic and quintic oscillators calculated
by means of the RPM.}
\label{tab:EnK3K5}
\begin{center}
\par
\begin{tabular}{|D{.}{.}{2}D{.}{.}{20}D{.}{.}{20}|}
\hline \multicolumn{1}{|c}{$n$}& \multicolumn{1}{c}{$\Re{E_n}$} &
\multicolumn{1}{c|}{$\Im{E_n}$}  \\
\hline

\multicolumn{3}{|c|}{$K=3$} \\

\hline

0 & 0.617160049538936737543 & -0.4483930225759328563 \\
1 & 2.1933097310211208676 & -1.5935327966748432597  \\
2 & 4.0363800198348283252 & -2.9326017436011248866 \\
3 & 6.03909710846479453   & -4.38766088005387693 \\
4 & 8.16189987482112373   & -5.92996736837917780  \\
5 & 10.3822957279796942   & -7.54317938470562561 \\

\hline

\multicolumn{3}{|c|}{$K=5$} \\

\hline

0 & 0.6396297978172518292 & -0.3080294760617796669 \\
1 & 2.396357680279750382  & -1.154025036407214392  \\
2 & 4.9177001900469598    & -2.3682395944315758    \\
3 & 7.91746214032848      & -3.812848812151728     \\
4 & 11.31798850540404     & -5.450456000157942    \\
5 & 15.062218927774       & -7.253582338538         \\

\hline
\end{tabular}
\par
\end{center}
\end{table}

\begin{table}[H]
\caption{Rayleigh-Ritz method for $H=p^2+ix^N$ with basis sets of $M$
harmonic-oscillator eigenfunction }
\label{tab:RR_ixN}
\begin{center}
\par
\begin{tabular}{|D{.}{.}{3}|D{.}{.}{20}|D{.}{.}{20}|}
\hline \multicolumn{1}{|c}{$M$} & \multicolumn{1}{|c|}{$N=5$} & \multicolumn{1}{c|}{$N=7$} \\
\hline

10  &  1.13770276661976  &  1.29785656512558   \\
20  &  1.16571028907153  &  1.22599499851804   \\
30  &  1.16477239347223  &  1.22470989807491   \\
40  &  1.16477042677832  &  1.22471162741409   \\
50  &  1.16477040815780  &  1.22471168644715   \\
60  &  1.16477040794314  &  1.22471168904864    \\
70  &  1.16477040794343  &  1.22471168965977    \\
80  &  1.16477040794342  &  1.22471168936849        \\
\hline
RPM &  1.16477040794341499419 &  1.2247116893311451 \\

 \hline
\end{tabular}
\end{center}
\end{table}

\begin{table}[H]
\caption{First eigenvalues of the PT-symmetric oscillators $H=p^2+ix^N$
calculated by means of the RPM, CRM and WKB method}
\label{tab:compa_ixN}
\begin{center}
\par
\begin{tabular}{|D{.}{.}{20}D{.}{.}{20}D{.}{.}{10}|}
\hline \multicolumn{1}{|c}{RPM}& \multicolumn{1}{c}{CRM} &  \multicolumn{1}{c|}{WKB}  \\
\hline

\multicolumn{3}{|c|}{$N=5$} \\

\hline

1.16477040794341499419 & 1.1647704079434150203   &   1.771244715  \\
4.3637843677121091602  & 4.3637843677121073149   &   8.509035978  \\
8.9551669982406716852  & 8.955166998240678966    &  17.65253759 \\

\hline

\multicolumn{3}{|c|}{$N=7$} \\

\hline

1.2247116893311451  &   1.2247116896597694535  &     2.855548625 \\
4.72146253539246    &   4.72144769127068       &    15.77168804  \\
10.0754495630818    &  10.0757623417291        &    34.91212093    \\

\hline
\end{tabular}
\par
\end{center}
\end{table}

\begin{table}[H]
\caption{First resonance for the quartic oscillator ($K=4$); in this case
the optimal rotation angle is $\theta_j=\frac{2j+1}{6}\pi$.}
\label{tab:E0j_K4}
\begin{center}
\par
\begin{tabular}{|D{.}{.}{2}|D{.}{.}{20}|D{.}{.}{20}|}
\hline \multicolumn{1}{|c}{$j$} & \multicolumn{1}{|c|}{$\Re\left
(E\right )$} & \multicolumn{1}{c|}{$\Im\left (E\right )$} \\
\hline
\multirow{2}{*}{0} & 0.33399312957789 & -0.57849306980780 \\
& 0.33399312957788855414 & -0.57849306980783854716 \\ \hline
\multirow{2}{*}{1} & -0.66798625915576 & 0.00000000000000 \\
& -0.66798625915577710827 & 0.00000000000000000000 \\ \hline
\multirow{2}{*}{2} & 0.3339931295779 & 0.57849306980787 \\
& 0.33399312957788855414 & 0.57849306980783854716 \\ \hline
\end{tabular}
\end{center}
\end{table}

\begin{table}[H]
\caption{Eigenvalues of the PT-symmetric oscillators (\ref{eq:H_PT}) for $N=4
$ and $N=5$}
\label{tab:En_N4_N5}
\begin{center}
\par
\begin{tabular}{|D{.}{.}{2}|D{.}{.}{20}|D{.}{.}{12}|}
\hline \multicolumn{1}{|c}{$n$} & \multicolumn{1}{|c|}{$N=4$} & \multicolumn{1}{c|}{$N=5$} \\
\hline
0 & 1.4771497535779945721  & 1.9082645782   \\
1 & 6.0033860833082771515  & 8.58722083623  \\
2 & 11.802433595134781580  & 17.7108090118 \\
\hline
\end{tabular}
\end{center}
\end{table}

\end{document}